\documentclass[aps,showpacs]{revtex4}
\usepackage{graphicx}

\def\ss{\scriptscriptstyle }

\begin{document}
\title{Nonequilibrium carriers in an intrinsic graphene under interband photoexcitation}
\author{A. Satou$^{1,2}$}
\author{F.~T. Vasko$^{1,3}$}
\email{ftvasko@yahoo.com}
\author{V. Ryzhii$^{1,2}$}
\affiliation{$^{1}$~University of Aizu, Ikki-machi, Aizu-Wakamatsu 965-8580, Japan \\
$^{2}$~Japan Science and Technology Agency, CREST, Tokyo 107-0075, Japan\\
$^{3}$~Institute of Semiconductor Physics, NAS of Ukraine, Pr. Nauki 41, Kiev, 03028,
Ukraine}
\date{\today}

\begin{abstract}
We study nonequilibrium carriers (electrons and holes) in an intrinsic
graphene at low temperatures under far- and mid-infrared (IR) radiation
in a wide range of its intensities. The energy distributions of carriers
are calculated using a quasiclassic kinetic equation which accounts for
the energy relaxation due to acoustic phonons and the radiative
generation-recombination processes associated with  thermal
radiation and the carrier photoexcitation by incident radiation.
It is found that the nonequilibrium distributions are determined
by an interplay between weak energy relaxation on acoustic phonons and
 generation-recombination processes as well as by the effect of pumping
saturation. Due to the effect of saturation, the carrier distribution
functions can exhibit plateaus around the pumping region at elevated
intensities. As shown, at sufficiently strong mid-IR pumping, the population
inversion can occur below the pumping energy. The graphene dc
conductivity as a function of the pumping intensity exhibits a pronounced
nonlinearity with a sub-linear region at fairly low intensities and
a saturation at a strong pumping. However, an increase in the
pumping intensity in very wide range leads only to a modest increase in
the carrier concentration and, particularly, the dc conductivity. The
graphene conductivity at mid-IR irradiation exhibit strong
sensitivity to mechanisms of carrier momentum relaxation.
\end{abstract}


\maketitle
\section{Introduction}

The features of the dynamics of carriers (electrons and holes) in graphene
and the mechanisms of their relaxation \cite{1} result in the exceptional
properties of graphene and a new device prospects (see \cite{2} for review).
The studies of optical phenomena, including
the Raman scattering (see\cite{3} and references therein), ultrafast spectroscopy,
\cite{4,5} and magnetooptics, \cite{6} can be used to reveal both the energy
spectrum parameters and the mechanisms of carrier scattering. The gapless
energy spectrum of graphene, with the characteristic velocity
$v_W\simeq 10^8$ cm/s which corresponds to the neutrinolike bandstructure
(Weyl-Wallace model) \cite{7}, provides its nontrivial optical properties due
to effective interband transitions in the far- and mid-infrared (IR) spectral
regions. The linear response of epitaxial graphene \cite{8} and graphite \cite{9}
was measured and the pertinent calculations were performed in Ref.\cite{10}.
Recently,~\cite{11} a fairly low threashold of the nonlinear response under
far- or mid-IR excitation
have been found and the photoconductivity of an intrinsic graphene was
calculated for a low-pumping region. But an essentially nonlinear regime of
response was not calculated and no experimental data concerning nonlinear
properties of graphene in this spectral region are available. Thus, an
investigation of the nonlinear response of graphene is timely now.

In this paper, we consider the nonequilibrium energy distributions of the carriers
under interband photoexcitation by far- or mid-IR pumping in a wide range of the
intensities. Using the obtained distributions, we analyze the nonlinear dependencies
of the carrier concentration and the dc conductivity on the pumping intensity and
frequency at different temperatures. The dynamic conductivity of graphene photoexcited
by far- or mid-IR radiation is calculated as well.

As shown below, the energy distribution of carriers are determined by an interplay
between quasielastic energy relaxation, which is ineffective at low energies, and
generation-recombination processes. There is a marked increase in the carrier
population in the range of energies where the phonon and radiative mechanisms of
relaxation are virtually compensated. The interband absorption saturation also affects
the energy distribution of carriers. The graphene dc conductivity as a function of
the pumping intensity exhibits a pronounced nonlinearity with a sub-linear region
at fairly low intensities and a saturation at a strong pumping. It is also found
that in a certain energy range of the pumping power the real part of the interband
contribution to the dynamic conductivity becomes negative, i.e. the negative
absorption takes place.

The paper is organized in the following way. The model under consideration is
described in Sec. II. In Sec. III, we present the nonequilibrium carrier distribution
as function of pumping intensity for a few pumping frequencies. The results of
calculations of the photoconductivity and the dynamic conductivity, as a response
on a weak dc electric field or a probe high-frequency field, are discussed in Sec. IV.
The brief discussion of the assumptions used in calculations and
concluding remarks are given in the last section.

\section{Model}
In order to obtain the energy distributions of the carriers in intrinsic graphene  we
use the  quasiclassic kinetic equation derived and analytically analyzed  previously
\cite{11} for the case of low intensity pumping.  The kinetic equation under
consideration accounts for the energy relaxation due to scattering on
acoustic phonons, the radiative generation-recombination processes associated
with  thermal radiation, and the far- or mid-IR pumping. We disregard the inter-carrier
scattering since at low temperatures the carrier concentration can be small even
at relatively strong interband pumping. Since the scattering mechanisms in $c$-
and $v$-bands are symmetric, the electron and hole distributions in the intrinsic
material are identical and we consider below the carrier distribution function $f_p$.
Taking into account the abovementioned mechanisms, the kinetic equation
under consideration, which governs the distribution function
$f_p$, is presented in the following form:
\begin{equation}\label{eq1}
J_{LA}\{f_p\} + J_{R}\{f_p\} + G\{ f_p\} = 0 .
\end{equation}
Here the collision integrals $J_{LA}\{f_p\}$ and  $J_{R}\{f_p\}$ are associated
with the relaxation of carriers caused by the acoustic phonons and  the equilibrium
thermal radiation, respectively, the term $G\{ f_p\}$ describes the interband
carrier excitation. Since the interband transition due to the acoustic phonon
scattering are forbidden (the sound velocity is weak in comparison to $v_W$),
the concentration balance equation takes form
\begin{equation}\label{eq2}
\int_0^\infty dpp[ J_R\{ f_p\} +G\{ f_p\}] = 0
\end{equation}
and it can be considered as the normalization condition for $f_p$. Another
condition for $f_p$ streams from the requirement $f_{p\rightarrow\infty}$=0, so that
any term in Eq.~(1) is equal to zero at $p\rightarrow\infty$, i.e. the zero acoustic
flow at high energies takes place, $J_{LA}\{f_p\}|_{p\rightarrow\infty}$=0.

Considering the quasielastic scattering of carriers on acoustic phonons,
one can use the Fokker-Planck form of $J_{LA}\{f_p\}$ \cite{12} and
Eq. (1) can be presented as~\cite{11}
\begin{eqnarray}
\frac{\gamma}{p}\frac{d}{dp}\left\{ p^4\left[\frac{df_p}
{dp}+\frac{f_p (1 -f_p )}{p_T}\right]\right\}   \nonumber \\
+ \frac{p}{p_T}\left[ N_{2p/p_T}(1-2f_p)-f_p^2\right]  \\
+ G(1-2f_p)\Delta\left(\frac{p - p_{\Omega}}{\delta p_{\Omega}}\right) = 0.  \nonumber
\end{eqnarray}
Here $p_T=T/v_W$ is the characteristic thermal momentum ($T$ is the
temperature in the energy units), $N_{2p/p_T}=[\exp (2p/p_T)-1]^{-1}$ is
the Planck distribution function, and the form-factor $\Delta(\varepsilon)$ describes
the broadening of interband transitions. So that $p_{\Omega} = \hbar\Omega/2v_W$
is the momentum of just photogenerated carriers  corresponding to the
pumping frequency $\Omega$ and the broadening is described by $\delta p_{\Omega}$
which is proportional to $p_{\Omega}$ if the scattering rates are proportional to the
density of states. It was assumed below that $\Delta(\varepsilon)= \pi^{-1/2}
\exp (- \varepsilon^2)$ and $\delta p_{\Omega}/p_{\Omega}\simeq$0.1 \cite{13}.
The dimensionless parameter $\gamma$ and $G$ are the relative strength of the energy
relaxation with respect to the generation-recombination efficiency and the
characteristic value of the pumping intensity, respectively. It is important below,
that a following dependencies on $T$, $\Omega$ and the pumping intensity, $S$,
take place:
\begin{equation}
\gamma \propto T~,~~~~~G\propto S/(\Omega^3T) ,
\end{equation}
so that the acoustic contribution increases with $T$ while the excitation efficiency
decreases with $\Omega$ and $T$. The explicit expressions of $g$ and $G$ can be
found in Ref.~\cite{11}.
Using the typical parameters of graphene \cite{14}, at $T$=77 K one obtains
$\gamma\simeq$0.32. At the same temperature
at $S$=1W/cm$^2$ and $\hbar\Omega$=120 meV
one can get $G\simeq$2.8.

Numerical procedure for the problem described can be simplified if one takes
into account that the equilibrium distribution remains valid at $p\rightarrow 0$.
The condition $f_{p\rightarrow 0}$=1/2 (it is the Fermi function at zero energy)
can be explicitly derived considering that  $G\{ f_p\}$ vanishes at 
$p\rightarrow 0$ (if $p_{\Omega}\gg\delta p_{\Omega}$) and equations 
$J_{LA}\{f_p\} = 0$ and $J_{R}\{f_p\} = 0$ are satisfied by the equilibrium distribution. Taking this into account, one can write the boundary condition at
$p\rightarrow\infty$, as the requirement that the factor $\{\ldots\}$ under 
derivative in the first term in the left-hand side of Eq.~(3) turns to zero.
As a consequence, for numerical solution of Eq.~(2), one can use
the following requirements as the boundary conditions for the distribution function:
\begin{equation}\label{eq5}
f_p\biggr|_{p\to 0} = \frac{1}{2}, \qquad  p^4\biggl[\frac{df_p}{dp} +
\frac{f_p(1 - f_p)}{p_T^2}\biggr]\biggr|_{p\to\infty} = 0 ,
\end{equation}
while the balance condition (2) should be used for checking of a numerical results
obtained. The numerical solution of Eq.~(3)
is performed below using  a finite difference method and the pertinent iteration
procedure.
\begin{figure}[ht]
\begin{center}
\includegraphics{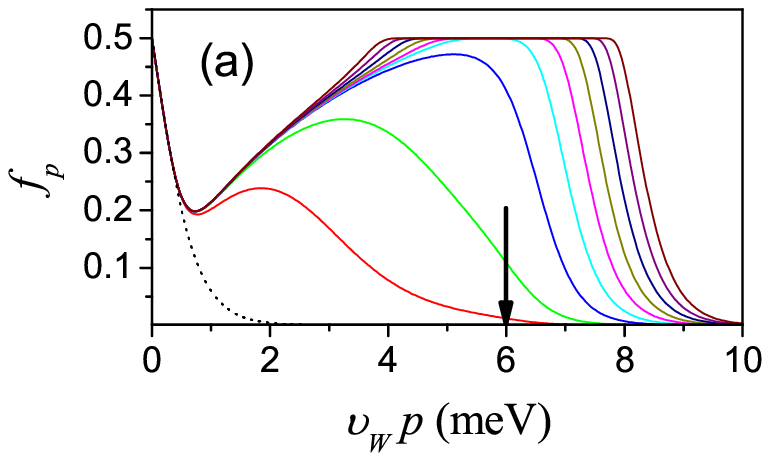}
\includegraphics{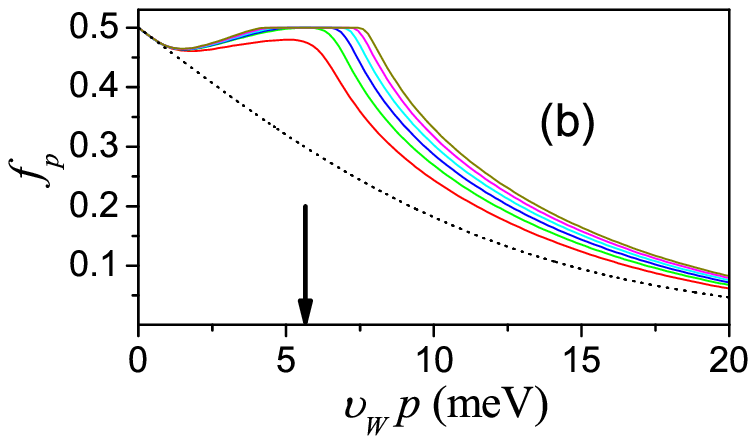}
\end{center}
\addvspace{-0.8 cm}\caption{(Color online)
Distribution functions $f_p$ vs energy $v_Wp$ (a) for the pumping intensities $S$=10$^{-5}$,
10$^{-4}, \ldots$, 10$^3$ W/cm$^2$ (from left to right) at $T = 4.2$~K and (b) for
$S$=10$^{-2}$, 10$^{-1}, \ldots$, 10$^3$ W/cm$^2$ at 77 K. Dotted curves correspond to
the equilibrium distributions. Vertical arrow corresponds to the energy equal to
$\hbar\Omega/2$ at $\hbar\Omega$=12 meV.}
\end{figure}

\section{Nonequilibrium distribution}
We present here the results of numerical solution of Eq.~(3) with
conditions (5) and  discuss the obtained distribution functions
at different   excitation conditions (frequency
and intensity of pumping) and temperature for the typical parameters of graphene
\cite{14}. The variation of  the sheet carrier
concentration  with varying excitation conditions is also considered.

\begin{figure}[ht]
\begin{center}
\includegraphics{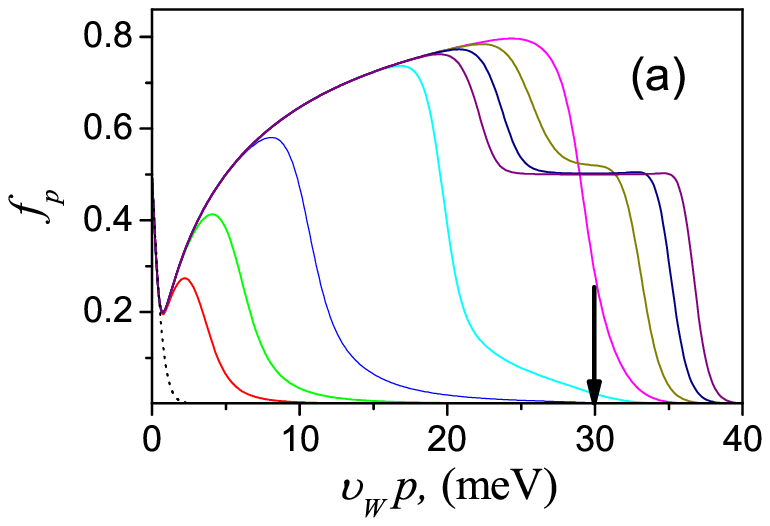}
\includegraphics{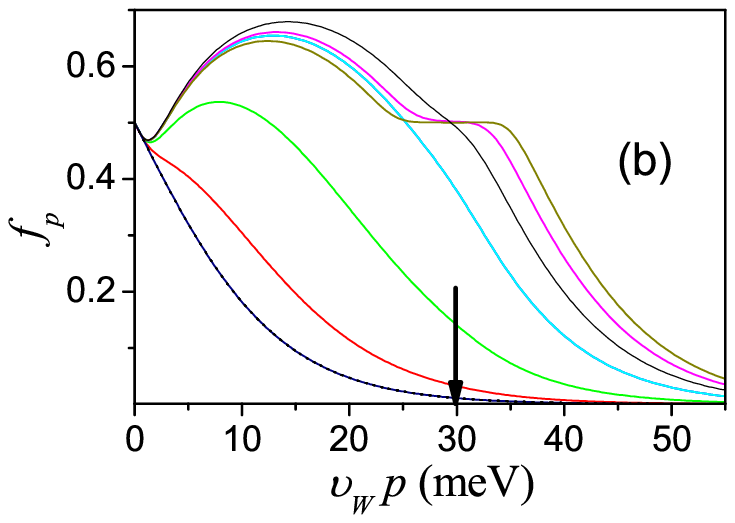}
\end{center}
\addvspace{-1 cm}\caption{(Color online) The same as in Fig. 1 under mid-IR pumping
with $\hbar\Omega =$60 meV.}
\end{figure}
\begin{figure}[ht]
\begin{center}
\includegraphics{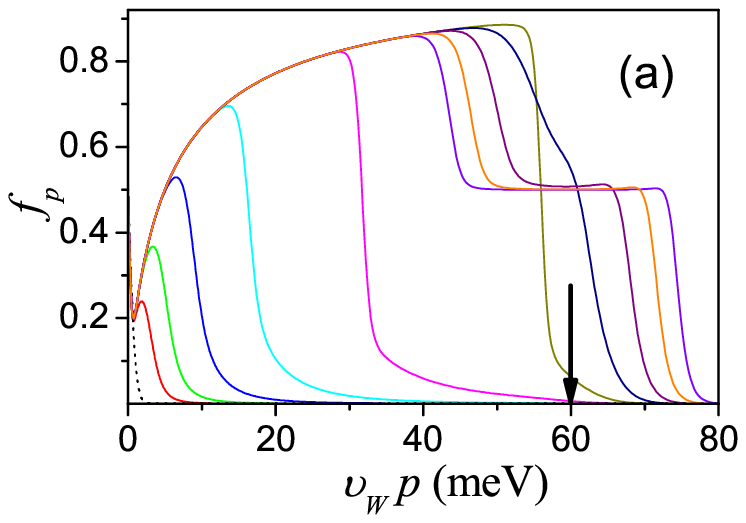}
\includegraphics{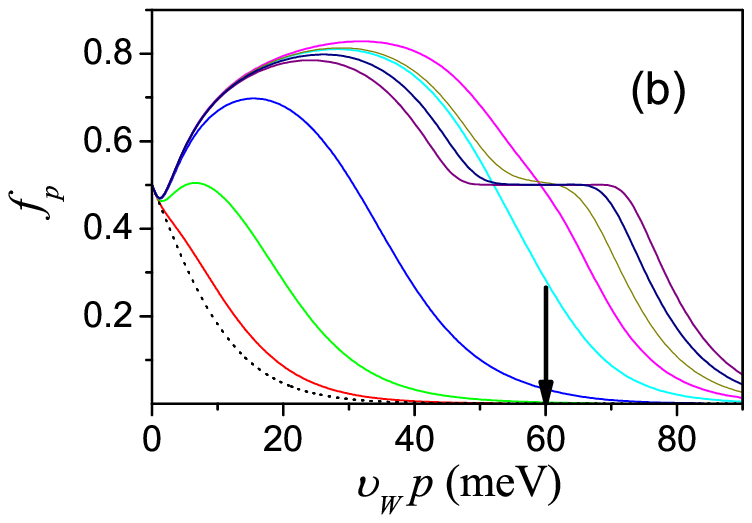}
\end{center}
\addvspace{-1 cm}
\caption{(Color online) The same as in Fig. 1 under the CO$_2$ laser pumping,
$\hbar\Omega =$120 meV, with intensities (a) $S$=10$^{-4}$, 10$^{-3}, \ldots$, 10$^5$
W/cm$^2$ (from left to right) and (b) for $S$=10$^{-2}$, 10$^{-1}, \ldots$, 10$^5$
W/cm$^2$.}
\end{figure}

The obtained distribution functions under the far-IR pumping with the photon 
energy $\hbar\Omega =$12 meV are shown in Fig. 1 with the one-order step in 
pumping intensities from $10^{-5}$ W/cm$^2$ to $10^3$ W/cm$^2$. First of all, 
one can see a visible modification of the carrier distributions at low intensities 
(up to  $S\sim 10^{-3}$ W/cm$^2$ at $T$=4.2 K and for $S\leq 10^{-1}$ W/cm$^2$ 
at $T$=77 K). Next, the peak of the distribution function shifts towards 
$\hbar\Omega /2$. Such a behavior is in agreement with the previous analytical consideration~\cite{11}. Further, at sufficiently strong pumping,
$f_{p \sim p_{\Omega}}$ tends also to the values close to $1/2$, so that plateau-like
energy distibutions are formed. The range of carrier energies,
where $f_p \simeq 1/2$, widens with increasing pumping intensity starting
$S\sim 10^{-2}$ W/cm$^2$ at $T$=4.2 K and $S\sim 0.1$ W/cm$^2$ at  $T$=77 K.
Since $f_p$ remains equilibrium at $p\to 0$ (all terms of Eq.(1) are equal zero
separately), one can see a deepening of $f_p$ in the region $v_Wp\sim T$.

Under mid-IR excitation, $\hbar\Omega =$60 meV, a similar character of the low-pumping
regime of response takes place, see Fig.~2.  With increased pumping intensity,
a peak of distribution appears, moreover, the peak distribution function can markedly
exceed $1/2$, i.e. the population inversion occurs if $S>10^{-2}$ W/cm$^2$
at $T$=4.2 K and $S>10^{-1}$ W/cm$^2$ at $T$=77 K. Further, in the pumping region
when $S>1$ W/cm$^2$ at $T$=4.2 K (or $S>10$ W/cm$^2$ at $T$=77 K)
$f_{p \sim p_{\Omega}}$ tends to $1/2$, so that a plateau of $f_p$ is formed 
around the energy
$\hbar\Omega /2$. As $S$ increases and the plateau region widens, the  peak
amplitude below $\hbar\Omega /2$ somewhat decreases but a maximal value of
$f_p$ exceeds 1/2. Since the equilibrium distribution is predetermined by
Eqs. (3) and (5) at $p\to 0$, a non-monotonic distribution with a deepening
at $v_Wp\leq T$, a peak under the energy $\hbar\Omega /2$, and a plateau around
$\hbar\Omega /2$ is realized.

Similar character of distribution takes place as $\hbar\Omega$ increases: the
same peculiarities are shifted to higher intensities according to Eq. (5). In
Figs. 3a and 3b we plot $f_p$ under the CO$_2$ laser pumping, $\hbar\Omega =$120 meV.
One can see that the population inversion regime begins for $S>10^{-2}$ W/cm$^2$
at $T$=4.2 K (or for $S>10^{-1}$ W/cm$^2$ at $T$=77 K) and the saturation region
around $\hbar\Omega /2$ takes place if $S>10^3$ W/cm$^2$ at $T$=4.2 K and 77 K.
These peculiarities are retained for higher intensities, up to $S\sim 10^5$ W/cm$^2$.

\begin{figure}[ht]
\begin{center}
\includegraphics{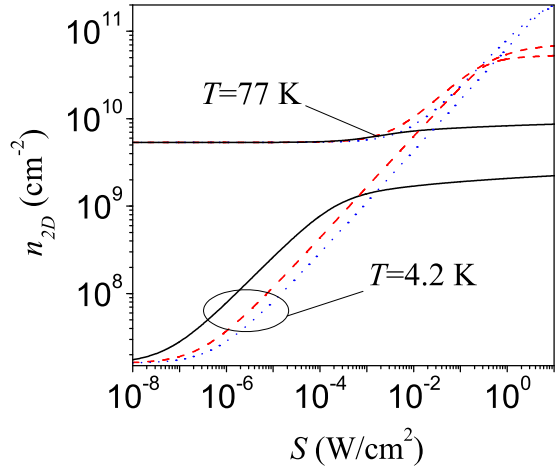}
\end{center}
\addvspace{-1 cm}
\caption{(Color online) Carrier concentration $n_{2D}$ vs  pumping intensity $S$
for the excitation energies $\hbar\Omega =$12, 60, and 120 meV (solid, dashed,
and dotted curves, respectively) at different temperatures.}
\end{figure}

\begin{figure}[ht]
\begin{center}
\includegraphics{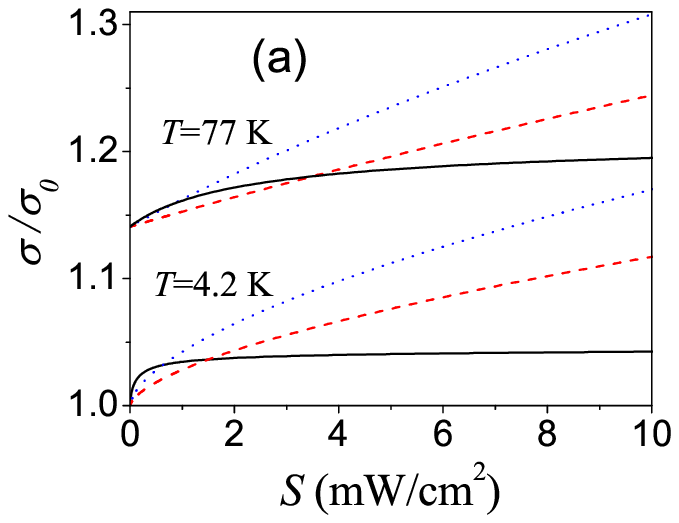}
\includegraphics{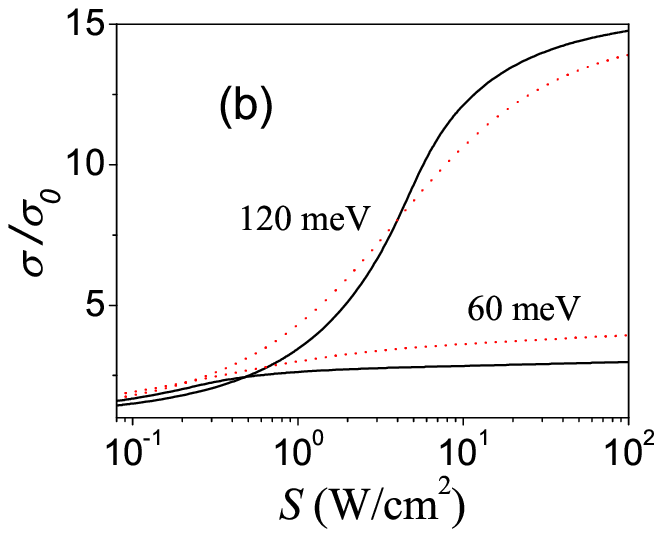}
\includegraphics{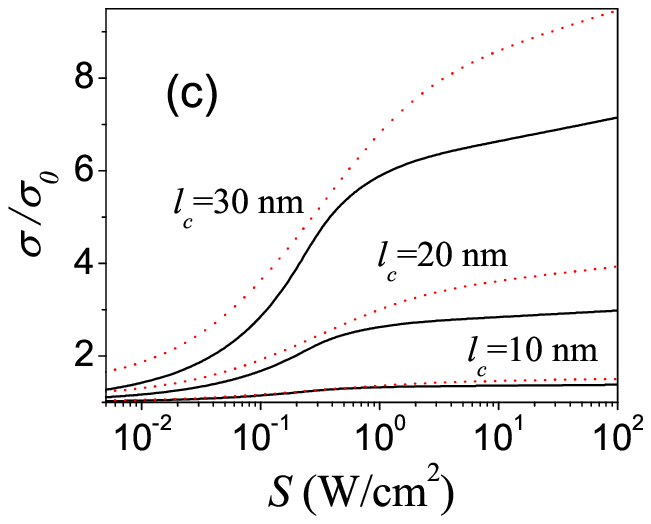}
\end{center}
\addvspace{-1 cm}
\caption{(Color online) Normalized dc conductivity $\sigma/\sigma_o$ vs  pumping
intensity $S$: (a) for low pumping region and $l_c$=20 nm at different temperatures
and $\hbar\Omega =$12, 60, and 120 meV (solid, dashed, and dotted curves, respectively),
(b) for mid-IR pumping at temperatures $T = 4.2$~K and 77~K (solid and dotted curves),
and (c) for different $l_c$ and $\hbar\Omega =$60 meV at $T = 4.2$~K and 77~K (solid
and dotted curves).}
\end{figure}

The obtained distribution functions allows to calculate the sheet carrier
concentration at different pumping conditions and temperatures according to the
standard formula:
\begin{equation}
n_{2D} = \frac{2}{\pi\hbar^2}\int_0^\infty dppf_{p} .
\end{equation}
Figure~4 demonstrates the dependences of the sheet concentration in graphene
as a function of the  pumping intensity corresponding to the energy distributions
shown in Figs.~1-3. As follows from the numerical calculation of the integral (6)
at $T$=4.2 K, $n_{2D}$ increases with $S$ in an intermediate pumping region and
tends to saturation under higher pumping intensities. The power dependence
$n_{2D}\propto S^r$ with $r\simeq$0.57 takes place at $T =4.2 K$ between
$S\sim 10^{-6}$ W/cm$^2$ and 0.5 W/cm$^2$ for the mid-IR pumping or between
$S\sim 10^{-7}$ W/cm$^2$ and $10^{-5}$ W/cm$^2$ for the far-IR pumping (see
Figs.~4a and 4b, respectively). Under mid-IR pumping at $T$=77 K, the concentration increases according to the same power law and saturates at the same pumping intensities. However, the nonlinear regime begins  starting  $S\sim 5$ mW/cm$^2$.
This is  because the equilibrium concentration is proportional to $T^2$. Note
that in the case of pumping by CO$_2$ laser, the saturation occurs at $n_{2D}
>10^{11}$ cm$^{-2}$. In this situation, the inter-carrier scattering might be
important.

\section{DC and dynamic conductivities}
Here we turn to the consideration of the response of
the nonequilibrium carriers
with  the distribution functions obtained above to a weak dc electric field
or a probe radiation. Taking into account  that the momentum relaxation of carriers
is caused by elastic scattering mechanisms, one can use the following formula
for the dc conductivity~$\sigma$~\cite{15}:
\begin{equation}
\sigma = \sigma_0\biggl[2f_{p=0}-\frac{l_c}{\hbar}
\int_0^{\infty}dpf_p\frac{\Psi '(pl_c/\hbar)}{\Psi (pl_c/\hbar)^2}\biggr].
\end{equation}
Here $l_c$ is the correlation length of static disorder scattering,
$\Psi (z)=e^{-z^2}I_1(z^2)/z^2$, where $I_1(z^2)$ the first order Bessel
function of imaginary argument, and $\sigma_0$ is the conductivity in the
case of short-range disorder scattering, when $l_c = 0$~\cite{15}.
According to Eq.(5) $f_{p=0} = 1/2$, for the short-range scattering case,
when $\overline{p}l_c/\hbar\ll 1$ ($\overline{p}$ is the average momentum),
the dc conductivity $\sigma$ is determined by the low energy carriers so
that $\sigma$ appears to be independent of optical pumping intensity
despite a significant concentration of the photogenerated carriers,
$\sigma\simeq\sigma_0$. For the definiteness, it was assumed that $l_c$=
10, 20, and 30~nm.

Figure~5 shows the dependences of the dc conductivity $\sigma$ normalized
by its characteristic value $\sigma_0$ on the pumping power $S$ calculated
using Eq.~(7) with the distribution functions obtained in Sec. III for $T = 4.2$~K
and $77$~K and different pumping energies $\hbar\Omega$. The dependences are
markedly nonlinear beginning from rather low pumping intensities (see the
pertinent curves in  Fig.~5a). The nonlinearity of the dc conductivity at
low $S$ ($\leq$1 mW/cm$^2$) is particularly strong in the case $T=4.2$~K and
far-IR pumping, $\hbar\Omega = 12$~meV, but the variation of conductivity
does not exceed $5\%$
and becomes saturated at $S>$5 mW/cm$^2$. In the range of
intermediate intensities of mid-IR pumping, see Fig. 5b, there is a power
increase in the dc conductivity, which is followed by the saturation region.
It is instructing that an increase in the pumping intensity by several
orders of magnitude leads to a modest increase in $\sigma /\sigma_0$.

In Fig.~5c, the normalized dc conductivity as a function of the pumping intensity
calculated for different values of the correlation length of the disorder
scattering, $l_c$. One can see that the conductivity becomes more sensitive to
the pumping when parameter $l_c$ increases. If $l_c$ tends to zero, a quenching
of photoconductivity takes place. Since the dc conductivity $\sigma$ depends on $\overline{p}l_c/\hbar$, it, as a function of the pumping intensity,
saturatates  at higher intensities that the carrier concentration $n_{2D}$
(compare Figs. 5 and 4).

\begin{figure}[ht]
\begin{center}
\includegraphics{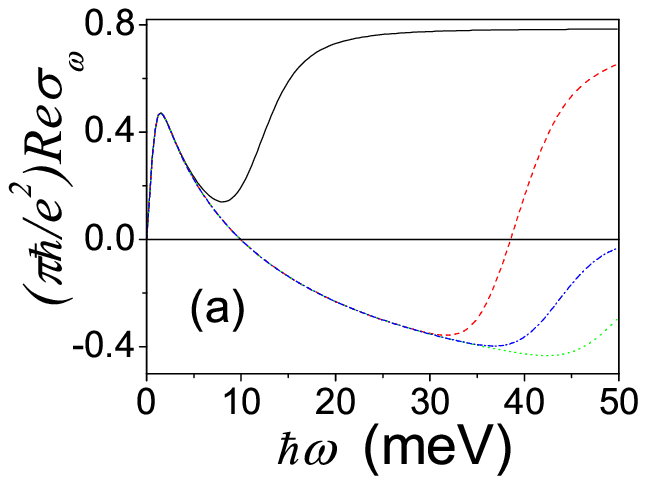}
\includegraphics{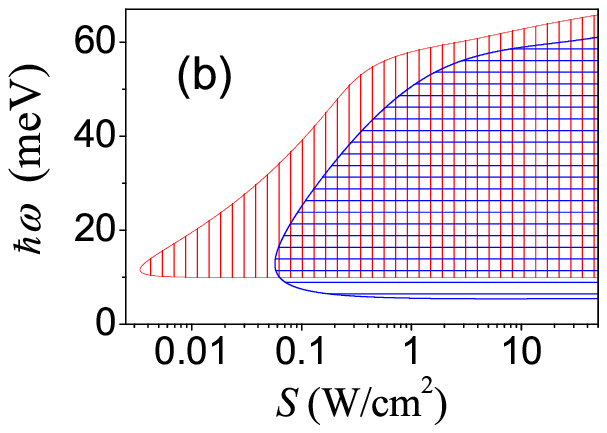}
\end{center}
\addvspace{-1 cm}\caption{(Color online) (a) Spectral dependences of the real
part of the dynamic interband conductivity $Re\sigma_{\omega}$ at $T = 4.2$~K
for $\hbar\Omega$=60 meV and pumping intensities: $S=10^{-3}$, 10$^{-1}$, 10$^1$,
and 10$^3$ W/cm$^2$ (solid, dashed, dotted and dash-dotted curves, respectively).
(b) Regions of parameters (shaded by vertical and horizontal lines for $T$=4.2 K
and 77 K, respectively) for which the dynamic interband conductivity is negative
if $\hbar\Omega$=60 meV.}
\end{figure}

As seen from Figs.~2 and 3, the distribution function $f_p$ can be larger than $1/2$
in a certain energy range if the pumping power is sufficiently high. This
corresponds to the population inversion of carriers and might be used for
far infrared  lasing. The real part of the interband contribution to the
dynamic conductivity which  is given by \cite{9}
\begin{equation}\label{eq7}
{\rm Re}\sigma_{\omega} =\frac{e^2v_W^2}{\hbar^2\omega}
\int_0^{\infty}dp\,p( 1 - 2f_{p})\,
\Delta\biggl(\frac{p - p_{\omega}}{\delta\,p_{\omega}}\biggr) ,
\end{equation}
can be negative. Here $p_{\omega}\equiv\hbar\omega/2v_W$ and $\delta p_{\omega}
\propto p_{\omega}$ are introduced in a similar way to $p_{\Omega}$ and
$\delta p_{\Omega}$ in Eq. (3).\cite{13}.  At $\delta p_{\omega}\to$0 one obtains
${\rm Re}\sigma_{\omega}<0$, if $f_{p_\omega}>1/2$. Figure~6a shows the
dependences of the real part of the interband
dynamic conductivity ${\rm Re}\sigma_{\omega}$ (normalized by factor $e^2/\pi\hbar$)
on the probe frequency $\omega$ at different pumping intensities $S$. In contrast
to the pumping scheme when the pumping energy $\hbar\Omega$ exceeds the energy
of optical photons, so that a cascade of optical phonons is emitted and the
photogenerated electrons (holes) are accumulated near the conduction band bottom
(valence band top), the negative dynamic interband conductivity in the case under
consideration corresponds to the interband transitions with relatively high
energies $\hbar\omega$ (compare with~\cite{16}). Figure~6 shows the regions
(shaded) on the pumping intensity - probe frequency plane corresponding to the
negative dynamic interband conductivity at different temperatures.
The dependence of the real part of net dynamic conductivity
on the probe frequency $\omega$, can be markedly modified
by the contribution of the intraband transitions (which correspond to the Drude
conductivity). However, the latter contribution can be effective only in the range
of rather small frequencies $\omega$.\cite{16}

\section{Discussion and Conclusions}

The main restriction the model used above is the neglect of the Coulomb
interaction of carriers, so that the obtained results are valid at their
sufficiently low concentrations. As can be found from Fig. 4, in the case
of an intrinsic graphene, this assumption limits the validity of our model
by the temperatures $T \lesssim 100 - 200$~K, whereas the limitation imposed
on the  pumping intensity is fairly liberal. The latter is owing to the effect
of saturation of the interband absorption associated with the Pauli principle.
In our consideration we also disregarded more complex processes, namely,
the carrier interactions with the substrate vibrations and the effect of
disorder on the generation-recombination processes. The mechanism included
in the model describe the general features of the relaxation processes
considered, so that their refinement should not lead to a qualitative
change of the pattern of the phenomena studied above. As for the photogeneration
of carriers by incident far- or mid-infrared radiation, only the single-photon
interband processes were  taken into account. Hence,  possible nonlinear
frequency multiplication and  renormalization of the energy spectra were
disregarded. Our estimates show that the latter correspond to the limitations
of the pumping intensity by the values $S < 0.5$~MW/cm$^2$
(at $\hbar\Omega = 12$~meV) and $5$~MW/cm$^2$ (at  $\hbar\Omega = 120$~meV).
The other assumptions of our model (the isotropic energy spectra of carriers,
the valley degeneration, and exclusion of the interaction with optical phonons
in the range of energy under consideration) appears to be rather natural.

In summary, we calculated the energy distributions of carriers in an intrinsic
graphene at low temperatures under far- and mid-IR ($\hbar\Omega =12 -120$~meV)
pumping in a wide range of its intensities (from zero to $10^2$~kW/cm$^2$).
It was shown that these distributions are determined by interplay
between weak energy relaxation on acoustic phonons and radiative
generation-recombination processes as well as by the effect of pumping
saturation due to the Pauli principle. The obtained energy distributions at
elevated pumping intensities can exhibit the plateau regions and the regions
corresponding to the population inversion. The graphene dc conductivity as a
function of the pumping intensity demonstrates a pronounced nonlinearity at
fairly weak pumping (particularly at helium temperature), a steep increase in
a certain  range of intermediate intensities, and a saturation at sufficiently
strong pumping. Due to this, the graphene dc conductivity varies through 1-2
orders of magnitude when the pumping intensity is varied many orders.
We showed that the alteration of the graphene dc conductivity with far- and
mid-IR irradiation, i.e., the effect of the photoconductivity is sensitive to
the correlation length of disorder scattering.

To conclude, the obtained results, can be useful for the extraction of the parameters
determined the relaxation mechanisms in graphene and for the estimation of a
potential of novel optoelectronic devices~\cite{16,17,18}, in particular, in
terahertz and far infrared lasers~\cite{16,17}.

\section*{Acknowledgments}
The authors (A.~S. and V.~R.) are grateful to Professor T.~Otsuji for stimulating
discussions. This work was partially supported by the Japan Science and Technology
Agency, CREST, Japan.



\begin{thebibliography}{52}
\bibitem{13}
One can estimate the broadening energy as $v_dp_\Omega$, so that
$\delta p/p_\Omega\simeq v_d/v_{\ss W}$. Here the characteristic
velocity $v_d$ was introduced in Ref. 15 through the momentum
relaxation time for the short-range scattering mechanism,
$\hbar /v_dp$.
\bibitem{14}
The characteristic values of the energy relaxation and pumping,
$\gamma$ and $G$, are introduced in Eq. (3) with the use of the energy
relaxation, the generation-recombination, and the photoexcitation rates
given by Eqs. (13), (10), and (7) of Ref. 11. Under numerical calculations
we used the following parameyers of graphene: the deformation potential
12 eV, the sheet density 7$\cdot 10^{-8}$ g/cm$^2$, the sound velocity
7.3$\times 10^5$ cm/s and the dielectric permittivity $\sim 3.7$ (SiO$_2$ -
graphene structure).
\end{thebibliography}
\end{document}